\documentclass[10pt,preprint]{aastex}

\usepackage{lscape}

\begin{document}


\title{Chemical analysis of Asymptotic Giant Branch stars in
  M62\footnotemark[1]} \footnotetext[1]{Based on observations
  collected at the ESO-VLT (Cerro Paranal, Chile) under program
  193.D-0232. Also based on observations (GO10120 and GO11609) with
  the NASA/ESA Hubble Space Telescope, obtained at the Space Telescope
  Science Institute, which is operated by AURA, Inc., under NASA
  contract NAS 5-26555.}

\author{
E. Lapenna\altaffilmark{2},
A. Mucciarelli\altaffilmark{2},
F. R. Ferraro\altaffilmark{2},
L. Origlia\altaffilmark{3},
B. Lanzoni\altaffilmark{2},
D. Massari\altaffilmark{3,4},
E. Dalessandro\altaffilmark{2},
}

\affil{\altaffilmark{2} Dipartimento di Fisica e Astronomia,
  Universit\`a degli Studi di Bologna, Viale Berti Pichat 6/2,
  I--40127 Bologna, Italy}
\affil{\altaffilmark{3} INAF-Osservatorio Astronomico di Bologna, Via
  Ranzani, 1, I-40127 Bologna, Italy}
\affil{\altaffilmark{4} Kepteyn Astronomical Institute, University of
  Groningen, Landleven 12, 9747 AD Groningen, The Nethelands}
%


\begin{abstract}
  We have collected UVES-FLAMES high-resolution spectra for a
  sample of 6 asymptotic giant branch (AGB) and 13 red giant branch
  (RGB) stars in the Galactic globular cluster M62 (NGC6266). Here we
  present the detailed abundance analysis of iron, titanium, and
  light-elements (O, Na, Al and Mg).  For the majority (5 out 6) of
  the AGB targets we find that the abundances, of both iron and
  titanium, determined from neutral lines are significantly
  underestimated with respect to those obtained from ionized features,
  the latter being, instead, in agreement with those measured for the
  RGB targets. This is similar to recent findings in other clusters
  and may suggest the presence of Non-Local Thermodynamical
  Equilibrium (NLTE) effects.  In the O-Na, Al-Mg and Na-Al planes,
  the RGB stars show the typical correlations observed for globular
  cluster stars.  Instead, all the AGB targets are clumped in the
  regions where first generation stars are expected to lie, similarly
  to what recently found for the AGB population of NGC6752.  While
  the sodium and aluminum abundance could be underestimated as a
  consequence of the NLTE bias affecting iron and titanium, the used
  oxygen line does not suffer from the same effects and the lack of
  O-poor AGB stars therefore is solid.  We can thus conclude that none
  of the investigated AGB stars belong to the second stellar
  generation of M62.  We also find a RGB star with extremely high
  sodium abundance ([Na/Fe]$=+1.08$ dex).
\end{abstract}


\section{Introduction}
\label{intro}

For stars with initial masses lower than $8 M_\odot$, the asymptotic
giant branch (AGB) is the last evolutionary stage characterized by
thermonuclear reactions (in two shells surrounding an inert
carbon-oxygen nucleus).  The extended and cool atmospheres of AGB
stars are ideal environments for the formation of dust grains and
molecules. Moreover, the strong stellar winds developing during this
phase return to the interstellar medium most of the material processed
during the star life, thus playing a crucial role in the chemical
evolution of the Universe.  These stars could also be at the origin of
the self-enrichment processes that occurred in the early stages of
globular cluster (GC) evolution, polluting the gas with the ejecta of
high-temperature CNO-burning products \citep{dercole08} and thus
producing the observed chemical anomalies in light elements and the
well-established Na-O and Mg-Al anti-correlations \citep[see, e.g.,][]
{carretta09_giraffe, carretta09_uves, mucciarelli09, gratton12}.

In spite of their importance and although the high luminosities of
these stars can dominate the integrated light of a stellar population
\citep[e.g.,][]{renzini86, fe95, hoefner98, vanloon99, cioni03,
maraston05, mucciarelli06}, only a few works have been dedicated to
the detailed study of their chemical patterns, especially in GCs where
the attention has been focused in particular on CN \citep{mallia78,
norris81, briley93}, iron \citep{ivans99, ivans01, lapenna14,
mucciarelli15_3201}, sodium \citep{campbell13, johnson14} and
proton-capture elements \citep{worley09}.  One of the first systematic
chemical analysis of AGB stars in GCs has been performed in M5 by
\cite{ivans01}, who found a significant discrepancy between the iron
abundance derived from neutral and from single-ionized lines, in the
sense of systematically lower values of [FeI/H], with respect to
[FeII/H]. Very similar results, with differences up to $\sim 0.2$ dex,
have been recently found in a sample of AGB stars in 47 Tucanae
(47Tuc) and NGC 3201 \citep[][respectively]{lapenna14, mucciarelli15_3201}.
In all cases, the discrepancy cannot be
explained by measurement uncertainties or an incorrect derivation of
the atmospheric parameters, and it is not observed in red giant branch
(RGB) stars belonging to the same cluster and analyzed in a
homogeneous way.  A proposed explanation is that AGB stars suffer from
departures from local thermodynamic equilibrium (LTE) conditions,
driven by iron over-ionization, in their atmospheres. In fact, this is
expected to mainly affect (weaken) the neutral lines, while leaving
unaltered the ionized features of the same chemical species (see
\citealt{mashonkina11}).  Following \citet{thevenin99}, important NLTE
effects are indeed expected in metal-poor stars with low values of
gravity, mainly comparable to those typically observed near the
RGB-tip of GCs, and they should decrease for increasing metallicity.
However, most of the giants studied by \citet{ivans01},
\citet{lapenna14} and \citet{mucciarelli15_3201} are much fainter than
the RGB-tip, and 47Tuc is one of the most metal-rich GCs ([Fe/H]$\sim
-0.8$; \citealp{lapenna14}). Moreover, the available NLTE corrections
are essentially the same for stars with similar atmospheric parameters
and cannot therefore explain why such a discrepancy is observed for
AGB stars only and not even in all of them.  Indeed, more recent
results obtained in the metal-poor GC M22 show that also some RGB
stars have FeI abundances significantly lower than those determined
from ionized lines \citep{mucciarelli15_m22}, thus adding further
complexity to this puzzling situation.

In order to help understanding the origin and the magnitude of these
effects, detailed chemical analyses of giant stars in GCs with
different metallicities and different properties are crucial.  In this
work we discuss the case of M62, for which we recently obtained
high-resolution spectra for a sample of 19 RGB and AGB stars.  This
cluster is the tenth most luminous Galactic GC (M$_V = -9.18$,
\citealt{harris96}, 2010 edition), located near the Galactic bulge and
affected by high and differential reddening, with an average color
excess $E(B-V)= 0.47$ mag \citep{harris96}.  It shows an extended
horizontal branch (HB) and hosts a large population of millisecond
pulsars and X-ray binaries and several blue straggler stars
\citep{damico01, pooley03, beccari06}.  However, in spite of its
noticeable properties, only one study about its chemical composition
by means of high-resolution spectra has been performed to date
\citep{yong14}, and it is based on 7 giant stars.

The paper is structured as follows. In Section \ref{obs} we describe
the observations and the spectral analysis performed. In Section
\ref{resu} we present the results obtained for the iron, titanium and
light element abundances.  Section \ref{concl} is devoted to the
discussion and conclusions of the work.


\section{Observations and spectral analysis}
\label{obs}

We have observed a sample of 19 giant stars in the GC M62 by using the
UVES-FLAMES@VLT spectrograph \citep{pasquini00} within the Large
Program 193.D-0232 (PI: Ferraro).  The spectra have been acquired by
using the grating 580 Red Arm CD\#3, which provides a high spectral
resolution (R$\sim$40000) and a spectral coverage between 4800 and
$6800\rm\mathring{A}$. The 19 targets have been sampled by
  means of four different fiber configurations, in five pointings of
  30 min each (one configuration has been repeated twice), during the
  nights of 2014, April 16 and June 2, 3 and 19. In each
  configuration, one or two fibers have been used to sample the sky
  for background subtraction purposes. After careful visual
  inspection, only the (19) spectra with a signal-to-noise larger than
  50 have been kept in the analysis.
The spectra have been reduced by using the dedicated ESO
pipeline\footnote{http://www.eso.org/sci/software/pipelines/}
performing bias subtraction, flat-fielding, wavelength calibration,
spectral extraction and order merging. The sky background has been
subtracted from each individual spectrum.

\begin{figure}[]
\includegraphics[trim=0cm 0cm 0cm 0cm,clip=true,scale=.80,angle=0]{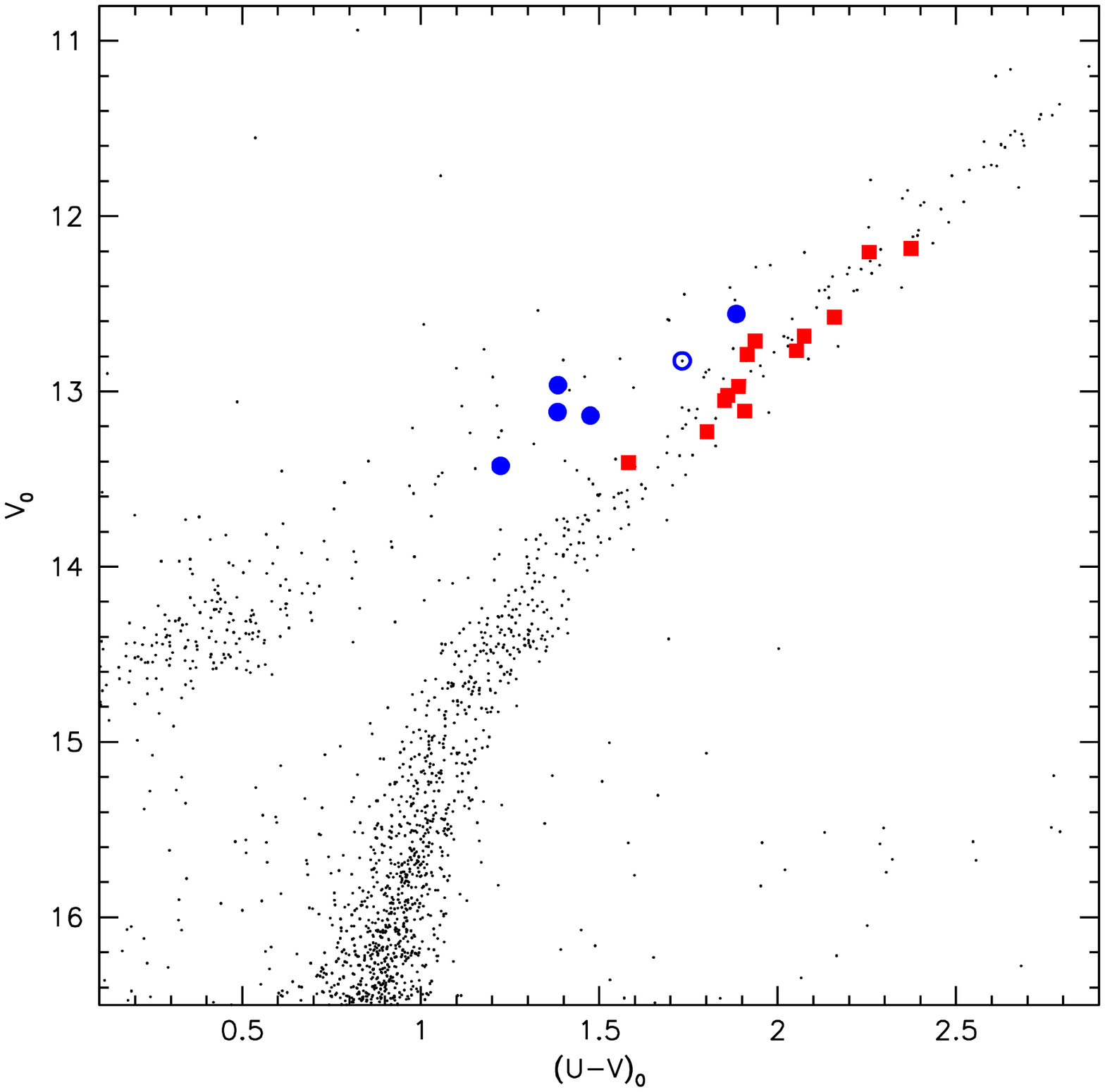}
\caption{Reddening-corrected color-magnitude diagram of M62, with the
  targets of the present study highlighted: 13 RGB stars (red squares)
  and 6 AGB objects (blue circles). The empty circle marks AGB star 96.}
\label{cmd}
\end{figure}

The target stars have been selected from the photometric catalog of
\citet{beccari06}, obtained from HST-WFPC2 observations. Only
stars brighter than $V=15$ and sufficiently isolated (i.e., with no
stellar sources of comparable or larger luminosity within a distance
of $2\arcsec$, and with no fainter stars within $1\arcsec$) have
been selected.  Figure \ref{cmd} shows the $(V, U-V)$
color-magnitude diagram (CMD) corrected for differential reddening
following the procedure described in \cite{massari12} and adopting the
extinction law by \cite{mccall04}.  The final sample includes 6 AGB
and 13 RGB stars.  All the target stars are located within
$\sim$85$\arcsec$ from the cluster center.  Their identification
number, coordinates, and magnitudes are listed in Table \ref{tab1}.

\subsection{Radial velocities}
\label{radial}
The radial velocities of our targets have been obtained by using the
code DAOSPEC \citep{stetson08} and by measuring the position of over
300 metallic lines distributed along the whole spectral range covered
by the 580 Red Arm of UVES-FLAMES. The uncertainties have been
computed as the dispersion of the velocities measured from each line
divided by the square root of the number of lines used, and they
turned out to be smaller than 0.05 km s$^{-1}$.  Finally, we applied
the heliocentric corrections computed with the IRAF task RVCORRECT.
For each spectrum, the zero-point of the wavelength calibration has
been accurately checked by means of a few emission lines of the
sky. The final velocities are listed in Table \ref{tab1}.  They range
from $-109.8$ km s$^{-1}$ to $-53.4$ km s$^{-1}$, with a mean value of
$-76.7 \pm 3.6$ km s$^{-1}$ and a dispersion $\sigma = 15.6$ km
s$^{-1}$. These values are in good agreement with the derivations of
\citet[][$v_r = -71.8 \pm 1.6$ km s$^{-1}$, $\sigma = 16.0$ km
  s$^{-1}$]{dubath97} and \citet[][$v_r = -70.1 \pm 1.4$ km s$^{-1}$,
  $\sigma = 14.3$ km s$^{-1}$]{yong14}, the small differences
  being likely due to the small statistics.

  The most discrepant target (id=79), with a radial velocity of
  $-109.85$ km s$^{-1}$, is still within 2$\sigma$ from the systemic
  velocity of the cluster. By using the Besan\c{c}on Galactic model
  \citep{robin03}, we extracted a sample of about 5300 field stars in
  the direction of M62, finding a quite broad and asymmetric radial
  velocity distribution, with mean $v_r \simeq -60$ km s$^{-1}$ and
  dispersion $\sigma = 80$ km s$^{-1}$, which partially overlaps with
  that of the cluster. On the other hand, only a few percent of the
  stars studied in that region close to the Galactic bulge have
  a [Fe/H] $< -$1.0 dex \citep[see e.g.][]{zoccali08,hill11,johnson13,ness13}.
  Thus, taking into account the metallicity of star 79 (see below), its position in the CMD,
  and its distance from the cluster center ($d \sim 38.5\arcsec$), we conclude
  that it is likely a genuine cluster member and we therefore keep it
  in the following analysis.

\subsection{Atmospheric parameters and stellar masses}
\label{atmos}
First guess effective temperature ($T_{\rm eff}$) and surface gravity
($\log g$) values for each target have been derived by using the
photometric information.  Temperatures have been estimated by using
the $(U-V)_{0}-T_{\rm eff}$ calibration of \cite{alonso99}.  Gravities
have been computed with the Stefan-Boltzmann equation by adopting the
color excess quoted above, a distance modulus $(m-M)_0 = 14.16$ mag
\citep{harris96} and the bolometric correction from \cite{alonso99}.
For the RGB stars we adopted a mass of 0.82 $M_{\odot}$, according to
the best fit isochrone retrieved from the PARSEC dataset
\citep{bressan12}, and computed for an age of 12 Gyr and a metallicity
Z=0.0013.  For the AGB stars we adopted a mass of 0.61 $M_{\odot}$,
according to the median value of the horizontal branch (HB) mass range
estimated by \cite{gratton10}.

Then we have performed a spectroscopic analysis as done in
\cite{lapenna14} and \cite{mucciarelli15_3201}, constraining the
atmospheric parameters as follows: (1) spectroscopic temperatures have
been obtained by requiring that no trend exists between iron abundance
and excitation potential, (2) the gravity was derived by using the
Stefan-Boltzmann equation with the value of $T_{\rm eff}$ thus
obtained and (3) the microturbulent velocity was determined by
requiring that no trend exists between iron abundance and line
strength.  In order to evaluate the effects of a different procedure
in the derivation of the atmospheric parameters and abundances, we
have also performed a spectroscopic determination of the surface
gravities by modifying condition (2) and imposing that the same
abundance is obtained from neutral and single-ionized iron lines (ionization balance).

\subsection{Chemical Abundances}
\label{chem}
The chemical abundances of Fe, Ti, Na, Al and Mg have been
derived with the package
GALA\footnote{http://www.cosmic-lab.eu/gala/gala.php}
\citep{mucciarelli13a}, which adopts the classical method to
  derive the abundances from the measured EWs of metallic unblended
  lines. The EW and the error of each line were obtained using
DAOSPEC, iteratively launched by means of the
4DAO\footnote{http://www.cosmic-lab.eu/4dao/4dao.php} code
\citep{mucciarelli13b}.  The lines considered in the analysis have
been selected from suitable synthetic spectra at the UVES-FLAMES
resolution and computed with the SYNTHE package \citep{sbordone05} by
using the guess atmospheric parameters and the metallicity derived by
\cite{yong14}.  The model atmospheres have been computed with the
ATLAS9\footnote{http://wwwuser.oats.inaf.it/castelli/sources/atlas9codes.html}
code. We adopted the atomic and molecular data from the last
release of the Kurucz/Castelli
compilation\footnote{http://wwwuser.oats.inaf.it/castelli/linelists.html}
and selected only the lines predicted to be unblended.
The selected lines and the atomic data adopted in the analysis are
listed in Table \ref{tab2}.

As detailed in Table \ref{tab3}, we used 100-150 FeI lines
and 7-12 FeII lines to derive the iron abundances, 25-60 lines of TiI
and 6-15 lines of TiII to derive the abundances of titanium.  For NaI,
MgI and AlI only few lines are available, namely those at
5682-5688$\rm\mathring{A}$ and 6154-6160$\rm\mathring{A}$ for NaI, the
line at 5711$\rm\mathring{A}$ and the doublet at
6318-6319$\rm\mathring{A}$ for MgI, and the doublet at
6696-6698$\rm\mathring{A}$ for AlI.  The O abundances have been
derived from spectral synthesis in order to take into account the
blending between the forbidden [OI] line at 6300.3$\rm\mathring{A}$
and a Ni transition.  For the Ni we adopted the average
  abundance obtained by \cite{yong14}, while for stars located in the
  upper-RGB we assumed average C and N abundances according to
  \cite{gratton00}, all rescaled to the assumed solar reference values
  \citep{grevesse98}. Because in some spectra the [OI] line was
partially blended also with a telluric line, the spectra have been
cleaned by using suitable synthetic spectra obtained with the TAPAS
tool \citep{bertaux14}.  For some stars, the [OI] line is not
detectable, thus only upper limits are obtained.  As solar reference
abundances we adopted the \cite{caffau11} value for O, and those of
\cite{grevesse98} for all the other elements.

For the computation of the global uncertainties on the final
abundances we took into account two main sources of errors, which have
been added in quadrature:
\begin{itemize}
\item[1)] the error arising from the EW measurements. For each star we
  computed this term by dividing the line-to-line dispersion by the
  square root of the number of lines used.  Thanks to the high-quality
  of the spectra and to the number of lines that can be exploited,
  this term turned out to be very small, especially for FeI and TiI
  (providing up to 150 lines).  For these species the line-to-line
  scatter is smaller than 0.1 dex, leading to internal uncertainties
  lower than 0.01-0.02 dex. For FeII and TiII the number of lines
  ranges from 7 up to 15, leading to an uncertainty of about 0.02-0.03
  dex. For the other chemical species the number of measured lines is
  much smaller (1-4).  Hence, the average uncertainties are of the
  order of 0.06-0.08 dex for OI, NaI, MgI and AlI.
\item[2)] the error arising from atmospheric parameters. For the
  computation of this term we varied each parameter by the $1\sigma$
  error obtained from the previous analysis. We have found that
  representative errors for $T_{\rm eff}$, $\log g$ and $v_{turb}$ are
  $\sim50$ K, $0.1$ dex and $0.1$ km s$^{-1}$, respectively, for both
  the RGB and the AGB samples.  Thus we decided to adopt these values
  as $1\sigma$ error for all stars. We also checked the effect
    of a $\pm 0.1$ dex change in the metallicity of the model
    atmosphere, finding variations smaller than $\pm 0.01$ dex on the
    final abundances.
\end{itemize}

\section{Results}
\label{resu}
The determination of abundances and abundance ratios of the various
chemical elements is described below. The adopted atmospheric
parameters and the measured iron and titanium abundances for the
observed RGB and AGB stars are listed in Table \ref{tab3}, while the
abundances of the light-elements are listed in Table \ref{tab4}.
In Table \ref{tab5} we present the global abundance uncertainty of one
RGB and one AGB star, as well as the uncertainties obtained by varying
each atmospheric parameter independently.  Since this approach does
not take into account the correlations among different parameters, the
global error can be slightly overestimated.

Since star 96 presents an anomalous behavior with respect to the
other AGB targets, in the following analysis it is not included in
the AGB sample (thus counting five stars), and it is discussed
separately at the end of Section \ref{sec:feti}.


\subsection{Iron and titanium}
\label{sec:feti}
By using spectroscopic gravities (thus imposing that the same iron
abundance is obtained from neutral and from single-ionized lines), the
average values measured for the RGB and the AGB sub-samples are
[Fe/H]$_{\rm RGB} = -1.10 \pm 0.01$ ($\sigma = 0.04$ dex) and
[Fe/H]$_{\rm AGB} = -1.18 \pm 0.01$ ($\sigma = 0.03$ dex).  These
values are consistent (within 1-2 $\sigma$) with previous abundance
determinations of M62 giants, regardless they are on the RGB or on the
AGB: [Fe/H]$=-1.12$ dex \citep{kraft03}\footnote{We refer to the
average value computed with Kurucz models without overshooting; see
\cite{kraft03} for details.}, [Fe/H] = $-1.18 \pm 0.07$ dex
\citep{carretta09_metscale}, and [Fe/H] = $-1.15 \pm 0.02$ dex
\citep[$\sigma = 0.05$ dex,][]{yong14}.

\begin{figure}[]
\includegraphics[trim=0cm 0cm 0cm
  0cm,clip=true,scale=.80,angle=0]{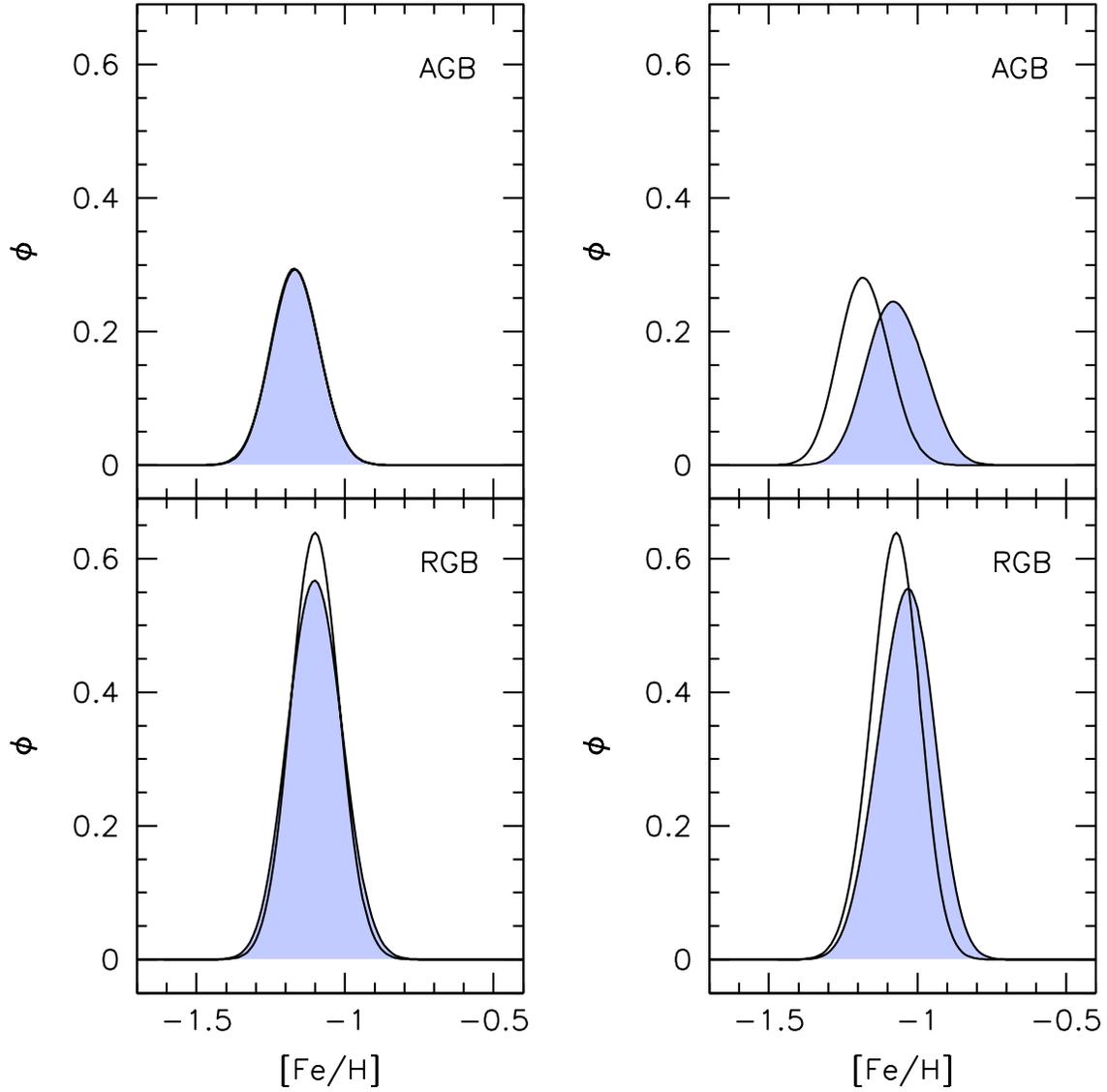}
\caption{\emph{Left panels:} generalized histograms for [FeI/H] (empty
  histograms) and [FeII/H] (blue shaded histograms) obtained by
  adopting spectroscopic gravities, for AGB stars (top panel) and the
  RGB sample (bottom panel). \emph{Right panels:} as in the left
  panels, but for the iron abundances obtained by adopting photometric
  gravities.}
\label{genhist}
\end{figure}

By using photometric gravities (and not imposing ionization balance),
we determined the iron abundances separately from neutral and from
single-ionized lines. For the 13 RGB stars we obtained [FeI/H]$_{\rm
  RGB} = -1.07 \pm 0.01$ dex ($\sigma = 0.04$ dex) and [FeII/H]$_{\rm
  RGB} = -1.04 \pm 0.02$ dex ($\sigma = 0.06$ dex).  For the 5 AGB
stars we measured [FeI/H]$_{\rm AGB} = -1.19 \pm 0.01$ dex ($\sigma =
0.04$ dex) and [FeII/H]$_{\rm AGB} = -1.06 \pm 0.02$ dex ($\sigma =
0.06$ dex). The average difference between the values of $\log g$
derived spectroscopically and those derived photometrically are
0.09 dex ($\sigma=0.10$ dex) and 0.30 dex ($\sigma=0.20$ dex) for the RGB and the
AGB samples, respectively. Figure \ref{genhist} shows the
generalized histograms of the iron abundances for the RGB and the AGB
samples separately, obtained by using spectroscopic (left panels) and
photometric gravities (right panels).  By construction, the
distributions of [FeI/H] and [FeII/H] essentially coincide if
spectroscopic gravities are assumed. Instead, the two distributions
significantly differ in the case of AGB stars if photometric gravities
are adopted. In particular, the average iron abundances of RGB stars
measured from neutral and single-ionized lines are consistent within
the uncertainties, while a difference of $-0.13$ dex, exceeding
5$\sigma$, is found for the AGB sample. Moreover, RGB and AGB stars
show very similar (well within 1$\sigma$) average values of [FeII/H],
while the neutral abundances of AGB stars are significantly lower (by
$0.12$ dex) than those of the RGB targets.

\begin{figure}[]
\includegraphics[trim=0cm 0cm 0cm
  0cm,clip=true,scale=.80,angle=0]{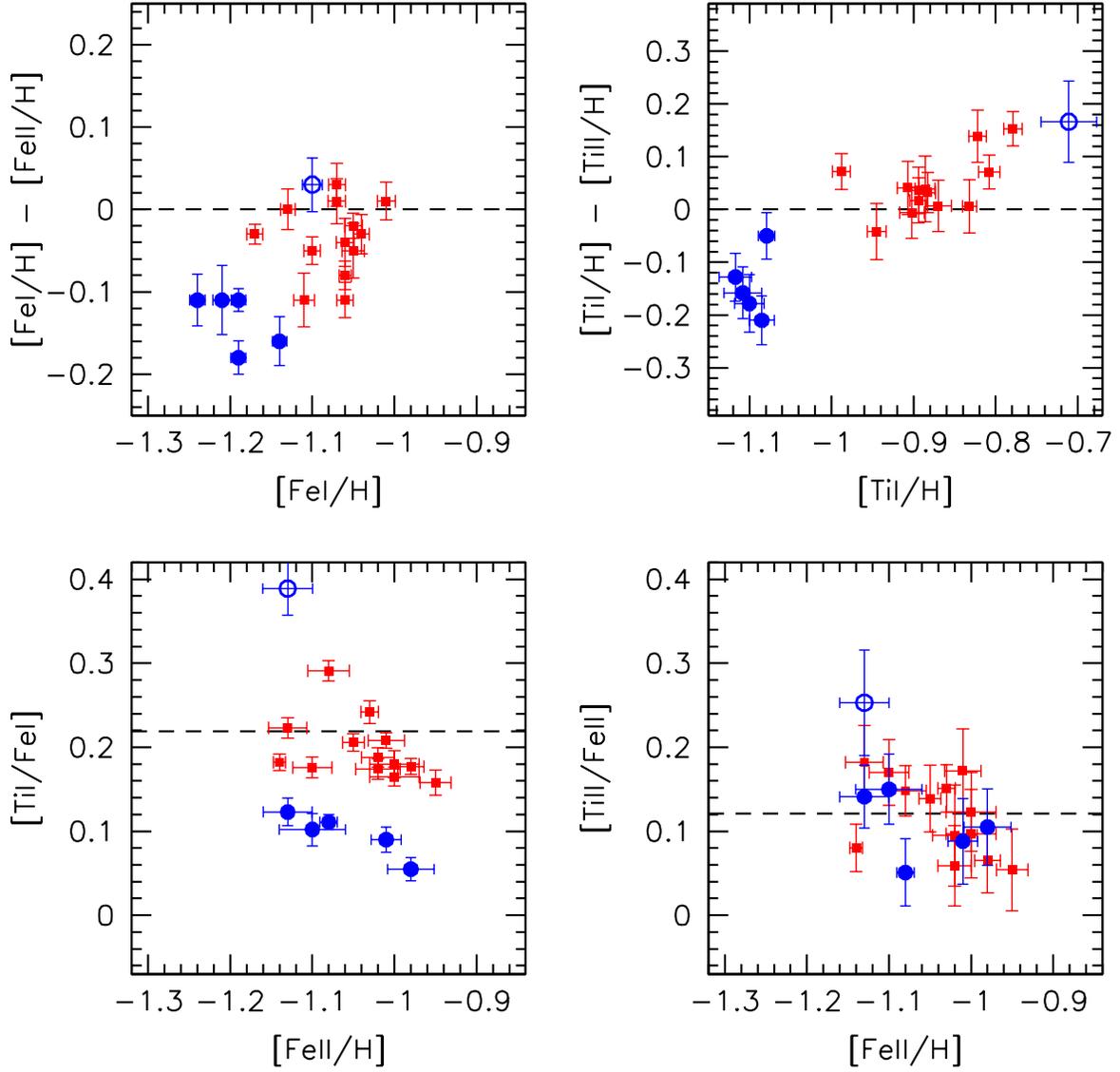}
\caption{\emph{Top panels:} difference between the chemical abundances
  derived from neutral and single ionized lines, as a function of that
  obtained from neutral lines, for iron (left panel) and titanium
  (right panel). Symbols are as in Fig. \ref{cmd}. \emph{Bottom
  panels:} [TiI/FeI] and [TiII/FeII] abundance ratios as a function
  of [FeII/H] for the studied samples.}
\label{feti}
\end{figure}

When using photometric gravities similar results are obtained
for titanium, the only other chemical species presenting a large
number of neutral and single-ionized lines.
For the RGB sample we find [TiI/H]$_{\rm RGB} =
-0.88 \pm 0.01$ dex ($\sigma = 0.06$ dex) and [TiII/H]$_{\rm RGB} =
-0.92 \pm 0.01$ dex ($\sigma = 0.05$ dex). For the AGB stars we
measure [TiI/H]$_{\rm AGB} = -1.10 \pm 0.01$ dex ($\sigma = 0.02$
dex) and [TiII/H]$_{\rm AGB} = -0.95 \pm 0.02$ dex ($\sigma = 0.06$
dex). In this case, the average abundance of AGB stars from neutral
lines is lower than that of the RGB sample by 0.21 dex (while such a
difference amounts to only 0.04 dex for the RGB sample).  In Figure
\ref{feti} we report the differences between the iron (top left panel)
and the titanium (top right panel) abundances derived from neutral and
from single-ionized lines, as a function of the abundances from the
neutral species, obtained for each observed star assuming photometric
gravities.  Clearly, with the only exception of star 96 (plotted as an
empty circle in the figure), the AGB and the RGB samples occupy
distinct regions in these planes, because of systematically lower
values of the AGB abundances derived from the neutral species.

\begin{figure}[]
\includegraphics[trim=0cm 0cm 0cm 0cm,clip=true,scale=.80,angle=0]{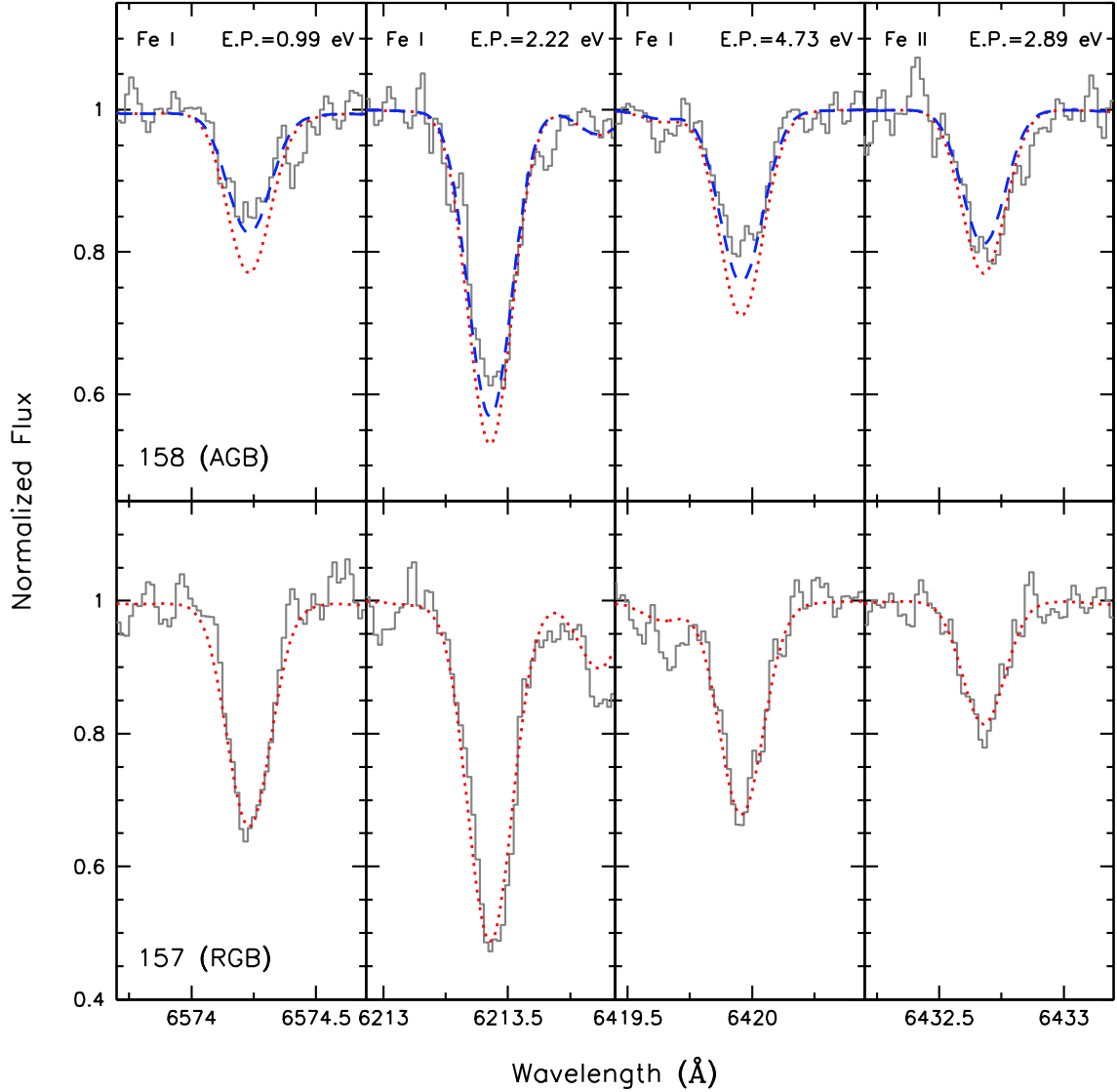}
\caption{Comparison between observed and synthetic spectra for AGB
  star 158 (upper panels) and RGB star 15 (lower panels), around three
  FeI lines with different excitation potentials and one FeII line
  (see labels).  The observed spectra are marked with gray lines.  The
  synthetic spectra have been computed by using the measured [FeI/H]
  (blue dashed line) and [FeII/H] (red dotted lines) abundances.
  Since the two abundances are practically identical for the RGB star,
  only one synthetic spectrum is shown in the lower panels.}
\label{spec}
\end{figure}

Such a difference can be also directly appreciated by visually
inspecting the line strengths in the observed spectra and their
synthetic best-fits.  As an example, in Figure \ref{spec} we show
the observed spectra of an RGB and an AGB star around some FeI and
FeII lines, together with synthetic spectra calculated with the
appropriate atmospheric parameters and the metallicity derived from
FeII and from FeI lines. As apparent, the synthetic spectrum
computed adopting the FeII abundance well reproduces all the
observed lines in the case of the RGB star, while it fails to fit
the neutral features observed in the AGB target, independently of
the excitation potential (thus guaranteeing that the effect cannot
be due to inadequacies in the adopted temperature). On the other
hand, the abundance measured from FeI lines is too low to properly
reproduce the depth of the ionized features of the AGB star.  This
clearly demonstrates a different behaviour of iron lines in AGB and
RGB stars.

To investigate the origin of the discrepancy between FeI and FeII 
abundances obtained for the AGB sample, we checked the impact of the 
adopted stellar mass on the estimate of the photometric gravity.  
As described in Sect. \ref{atmos}, for the AGB
stars we assumed a mass of $0.61 M_\odot$, corresponding to the median
value of the distribution obtained for HB stars by \cite{gratton10},
ranging from 0.51 to 0.67$M_{\odot}$.  By adopting the lowest mass
(0.51$M_{\odot}$), the average value of $\log g$ decreases by
$\sim0.08$ dex, while assuming the largest value, $\log g$ increases
by $0.04$ dex.
Such small gravity variations\footnote{Note that the
increase of $\log g$ is essentially the same ($0.05$ dex) even if
the mass provided by the best-fit isochrone ($0.72 M_\odot$) is adopted.}
have a negligible impact on the abundances derived from
the neutral iron lines, and the impact is still modest
(at a level of a few hundredths of a dex) on the abundances derived from
single-ionized lines. The only way to obtain (by construction) the
same abundance from FeI and FeII lines is to use the spectroscopic
values of $\log g$ derived from the ionization balance
(Sect. \ref{atmos}). However, these gravities correspond to stellar
masses in the range 0.25-0.3 $M_\odot$, which are totally unphysical
for evolved stars in GCs.

A possible explanation of the observed discrepancy could be a
departure from LTE condition in the atmosphere of AGB stars.
  In fact, lines originated by atoms in the minority ionization state
  usually suffer from NLTE effects, while those originated by atoms in
  the dominant ionization state are unaffected (see, e.g.,
  \citealt{mashonkina11}). Thus, if this is the case, the most
  reliable determination of the iron abundance is that provided by
  [FeII/H], since the majority of iron atoms is in the first ionized
  state in giant stars. Moreover, following \citet{ivans01}, the
degree of overionization of the neutral species should be (at least at
a first order) the same as the one affecting FeI lines.
Hence, the correct way to obtain a [X/Fe] abundance ratio is
  to compute it with respect to the FeI abundance if [X/H] is derived
  from minority species, and with respect to FeII if [X/H] is obtained
  from majority species. In the lower panels of Figure \ref{feti} we
present the [TiII/FeII] and the [TiI/FeI] abundance ratios as a
function of the iron abundance derived from single-ionized lines.

As expected, the abundances of AGB stars agree with those of the RGB
sample when single ionized (dominant state) titanium lines are
used. For [TiI/FeI] a systematic offset of the AGB sample toward
lower values is still observable (although reduced), thus indicating
the possible presence of residual NLTE effects. We also note
a systematic offset of +0.08 dex between [TiI/FeI] and [TiII/FeII],
especially for RGB stars.  However, taking into account that the
oscillator strength values of the TiII lines are highly uncertain and
that the offset is still reasonably small, we can conclude that the
[X/Fe] abundance ratio can be safely constrained either by neutral or
by single-ionized lines.  It is also interesting to note that the
average [TiI/Fe] and [TiII/Fe] abundance ratios (+0.16 dex and +0.25
dex, respectively) of \citet{yong14} show a relative offset of $-0.09$
dex, which is similar to ours but in the opposite direction. This
suggests that there is an intrinsic (although small) uncertainty in
the zero point scale of the titanium abundance.

\emph{AGB star 96 --} As apparent from Fig. \ref{feti}, AGB star 96
shows a difference between neutral and ionized abundances, both for
iron and titanium, which is incompatible with those found for the
other AGB targets, and which is much more similar to the values
measured for RGB stars. Interestingly, star 96 presents
atmospheric parameters compatible with those spanned by the RGB
targets (but with a surface gravity which is 0.15-0.2 lower than
that of RGB stars at the same temperature). This case is similar
to that encountered in 47Tuc, where the FeI abundance of a small
sub-sample of AGB stars (4 out of 24) has been found to agree with the
value obtained from ionized lines, thus suggesting that one of
the possible explanations could be the lack of LTE departures for
these objects \citep{lapenna14}. Also in M22, one (out of five)
AGB star shows a perfect agreement between [FeI/H] and [FeII/H], while
the other AGB stars show systematically low [FeI/H] values
\citep{mucciarelli15_m22}.


\subsection{Oxygen, sodium, magnesium and aluminum}
In most Galactic GCs, the abundances of oxygen, sodium, magnesium and
aluminium are known to show large star-to-star variations, organized
in clear correlations (see \citealt{gratton12} for a review).  These
are usually interpreted as the signature of self-enrichment processes
occurring in the early stages of GC evolution and giving rise to at
least two stellar generations with a very small (if any) age
difference, commonly labelled as first and second generations (FG and
SG, respectively).  In particular, the variations observed in O, Na,
Mg and Al are thought to be due to the ejecta from still unclear
polluters, like massive AGB stars, fast-rotating massive stars and/or
massive binaries \citep{fenner04, ventura05, decressin07, demink09,
marcolini09, bastian13, bastian15}.

\begin{figure}[]
\includegraphics[trim=0cm 0cm 0cm
  0cm,clip=true,scale=.80,angle=0]{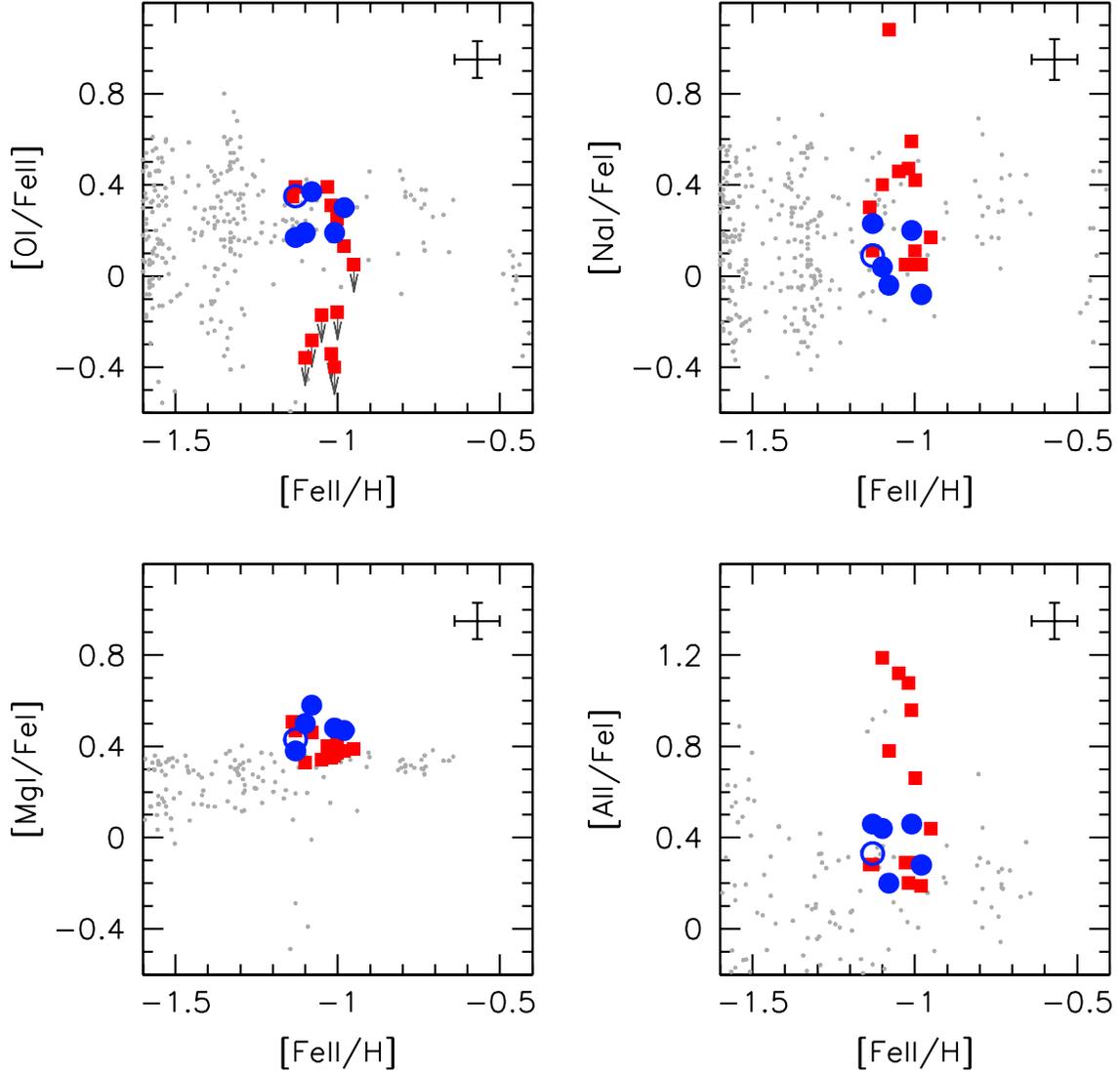}
\caption{From top-left to bottom-right, oxygen, sodium, magnesium and
  aluminum abundance ratios as a function of [FeII/H] for the studied
  sample of stars (same symbols as in Fig. \ref{cmd}). For a
  sub-sample of (O-poor) RGB stars, only upper limits to the oxygen
  abundance could be measured from the acquired spectra (see
  arrows). Representative error bars are marked in the top-right
  corner of each panel.  The values measured in large samples of giant
  stars in 20 Galactic GCs (from GIRAFFE and UVES spectra by
  \citealp{carretta09_giraffe, carretta09_uves, carretta14}),
  rescaled to the solar values adopted in this work, are shown for
  reference as grey dots.  }
\label{light}
\end{figure}

To verify the presence of these key features also in our sample of
giants, we derived the abundances of O, Na, Al, and Mg from the
observed spectra. The results are shown in Figure \ref{light}, where
all abundance ratios are plotted as a function of the iron content as
measured from the ionized lines. Since the oxygen abundance derived
from the forbidden [OI] line at 6300.3 $\rm\mathring{A}$ is not
affected by NLTE, its abundance ratio is expressed with respect to the
``true'' iron content (measured from FeII lines).  Instead, the other
species are known to suffer from NLTE effects and their abundances are
therefore plotted with respect to FeI (see Section
\ref{sec:feti}). This is true also for sodium, although we have
applied the NLTE corrections of \cite{gratton99}, which take into
account departures from LTE conditions driven by over-recombination
\citep{bergemann14}.\footnote{By adopting the NLTE corrections
of \citet{lind11}, the differential behaviour between AGB and RGB
stars remains the same.} In any case, we have verified that the
same results are obtained if the Na, Al and Mg abundances are computed
with respect to FeII or H. In agreement with what commonly observed in
Galactic GCs, we find that the Mg abundance is constant within the
uncertainties, while O, Na, and Al show significant (several tenths of
a dex) star-to-star variations in the RGB sample \citep[see
  also][]{yong14}. As shown in Figures \ref{ONa} and \ref{Al}, the
observed star-to-star variations are organized in the same
correlations observed for GCs.  In particular, oxygen and sodium are
anti-correlated, independently of using FeI or FeII for the
computation of the sodium abundance ratio (Fig. \ref{ONa}), while
aluminum and sodium are positively correlated and [AlI/FeI] shows a
$\sim 1$ dex spread for fixed magnesium (Fig. \ref{Al}).  Very
interestingly, instead, all abundance ratios are constant for the AGB
sample, with values mainly consistent with those commonly associated
to the FG.

The Na-O anti-correlation derived from our RGB sample is qualitatively
compatible with that measured by \citet{yong14}, who found two groups
of stars well-separated both in [Na/Fe] and [O/Fe]. We note that the
oxygen abundances quoted by \citet{yong14} are larger than ours, with
an average offset of +0.5 dex for the O-rich stars.  The origin of
this discrepancy can be ascribable to different factors (like atomic
data, telluric correction, etc.), but it is beyond the aims of this
paper. A good agreement with the results of \citet{yong14} is found
also for the [Al/Fe] and [Mg/Fe] distributions.

The derived Na-O anti-correlation of M62 is more extended than those
observed in most Galactic GCs.  Two discrete groups of stars can be
recognized, a first one with [O/Fe]$\sim$+0.2/+0.3 dex and
[Na/Fe]$\sim$+0.1 dex, and a second group with [O/Fe]$<$0.0 dex (only
upper limits) and [Na/Fe] at $\sim$+0.5 dex.  In particular, the
sub-solar O-poor component \citep[the so-called ``extreme
  population''; see ][]{carretta10_sgen} is quite prominent in M62,
while these stars are usually rare, observed only in some massive
systems, as NGC 2808 \citep{carretta09_giraffe}, M54
\citep{carretta10_m54}, and $\omega$ Centauri \citep{johnson10}.  We
also find a significant lack of ``intermediate population'' stars
\citep[with enhanced Na abundances and mild oxygen
  depletion][]{carretta10_sgen}, which are instead the dominant
component in most GCs.

\begin{figure}[]
\includegraphics[trim=0cm 0cm 0cm 0cm,clip=true,scale=.80,angle=0]{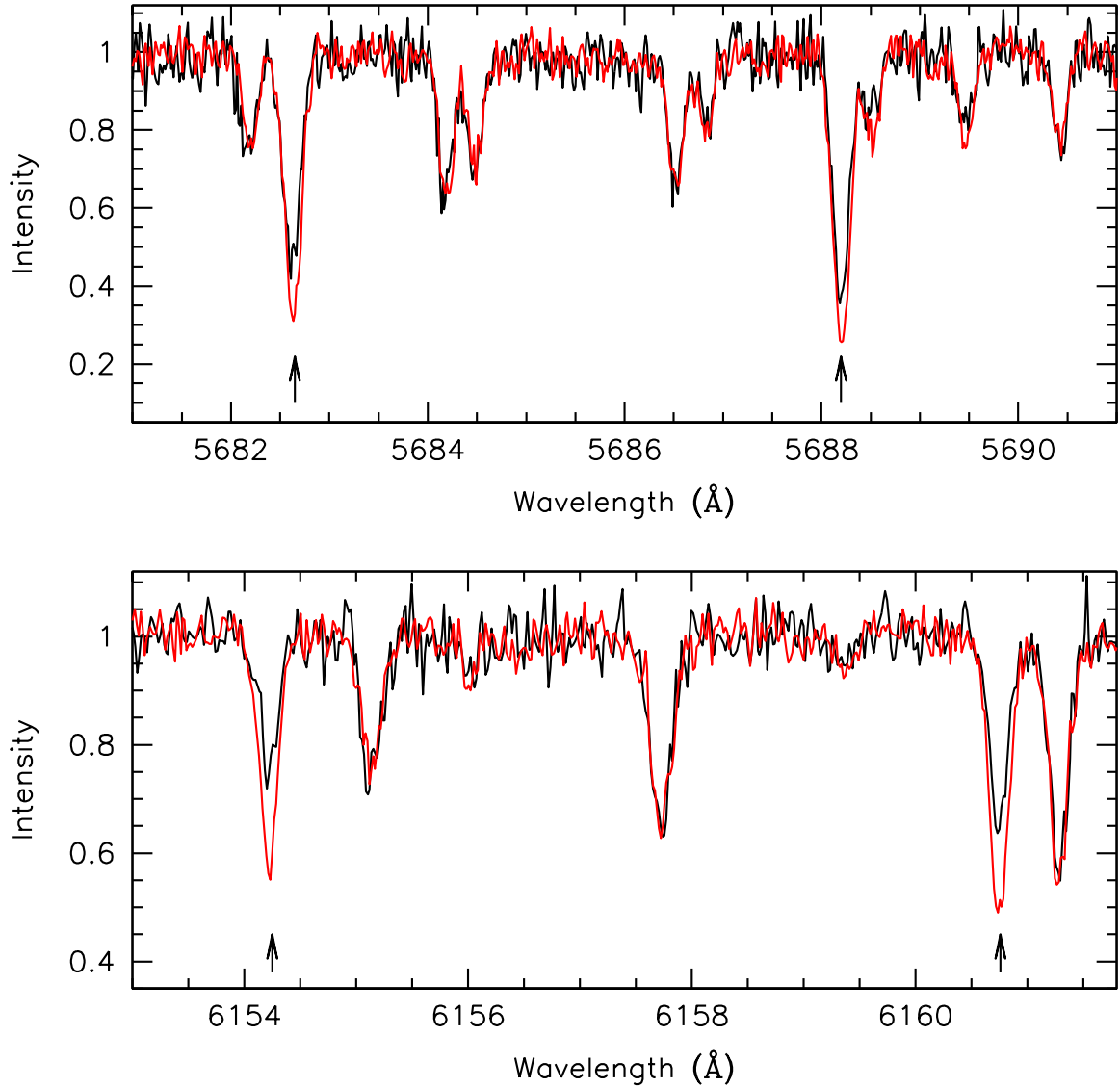}
\caption{Comparison between the spectra of the RGB stars 89 (red line)
  and 95 (black line) for the NaI lines at 5682-5688 $\rm\mathring{A}$
  (top-panel) and 6154-6160 $\rm\mathring{A}$ (bottom-panel).  The
  black arrows mark the position of the Na lines.}
\label{spec89}
\end{figure}

We finally note that the RGB star 89 exhibits a Na abundance
[Na/Fe]=+1.08 dex, which is $\sim$0.5 dex larger than that measured
for all the other O-poor stars.  In Figure \ref{spec89} we compare the
spectrum of star 89 with that of another RGB target (id=95) having
very similar atmospheric parameters and iron abundances (see Table
\ref{tab3}). As apparent, all lines have the same strengths, with the
notable exception of the two Na doublets, which are significantly
stronger in star 89.

To our knowledge, this is one of the most Na-rich giant ever
detected in a genuine GC (see also the comparison with literature
data in Fig. \ref{ONa}), with a [Na/Fe] abundance even higher than
the most Na-rich stars observed in NGC 2808 \citep{carretta06} and
NGC 4833 \citep{carretta14}, and comparable to a few extremely
Na-rich objects observed in the multi-iron system $\omega$ Centauri
\citep{marino11}.


\section{Discussion and conclusions}
\label{concl}
The differences in the iron and titanium abundances measured from
neutral and from single-ionized lines in five AGB stars of M62 closely
resemble those found in M5 \citep{ivans01}, 47Tuc \citep{lapenna14}
and NGC 3201 \citep{mucciarelli15_3201}.  These results might be
explained as the consequences of departures from LTE conditions,
affecting the neutral species, while leaving unaltered the ionized
lines.  The final effect is a systematic underestimate of the chemical
abundances if measured from neutral features.  Interestingly, the
findings in M5, 47Tuc and NGC 3201 seem to suggest that this effect
concerns most, but not all, AGB stars, while it is essentially
negligible for RGB targets.  This is inconsistent with the available
NLTE calculations \citep[e.g.][]{lind12, bergemann12}, which predict
the same corrections for stars with similar parameters. Moreover, the
results recently obtained in M22 show that the situation is even more
complex. In M22, in fact, neutral iron abundances systematically lower
than [FeII/H] have been measured also for RGB stars
\citep{mucciarelli15_m22}.  However, the [FeI/H]-[FeII/H] difference
in M22 clearly correlates with the abundance of s-process elements
(that show intrinsic star-to-star variations unusual for GCs),
suggesting that this could be a distinct, peculiar case. All
  these results seem to suggest that we are still missing some crucial
  detail about the behaviour of chemical abundances in the case of
  departures from LTE conditions, and/or that other ingredients (as
  full 3D, spherical hydro calculations, and the inclusion of
  chromospheric components) should be properly taken into account in
  modeling the atmospheres of these stars.

\begin{figure}[]
\includegraphics[trim=0cm 10cm 0cm
  0cm,clip=true,scale=.80,angle=0]{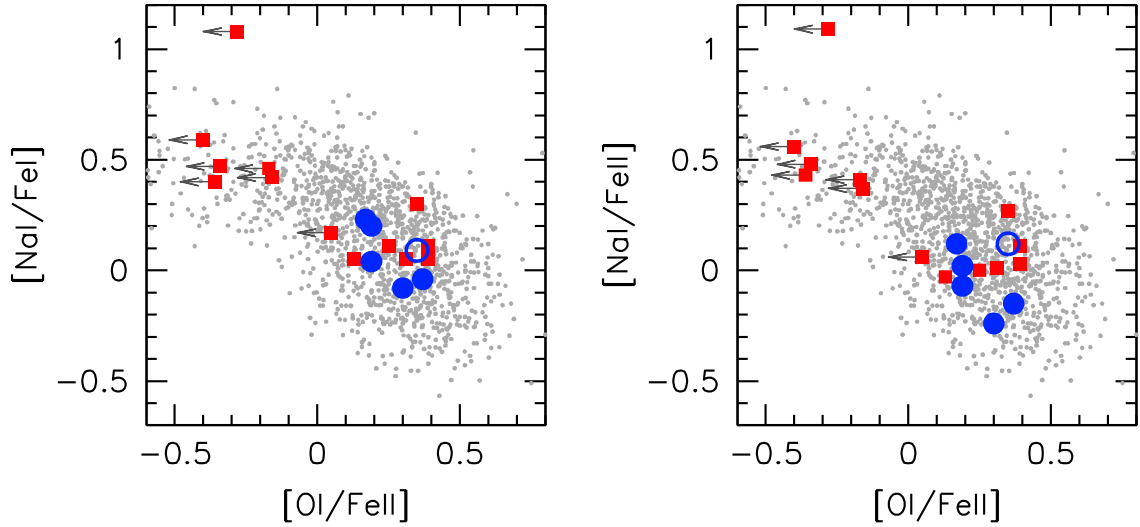}
\caption{Oxygen-sodium anti-correlation measured for the observed
  stars (same symbols as in Fig. \ref{light}). The corrections for
  NLTE effects provided by \citet{gratton99} have been applied to the
  Na abundances. This is then expressed with respect to FeI and to
  FeII in the left and right panels, respectively. Grey dots
  are as in Fig. \ref{light}.}
\label{ONa}
\end{figure}

\begin{figure}[]
\includegraphics[trim=0cm 10cm 0cm 0cm,clip=true,scale=.80,angle=0]{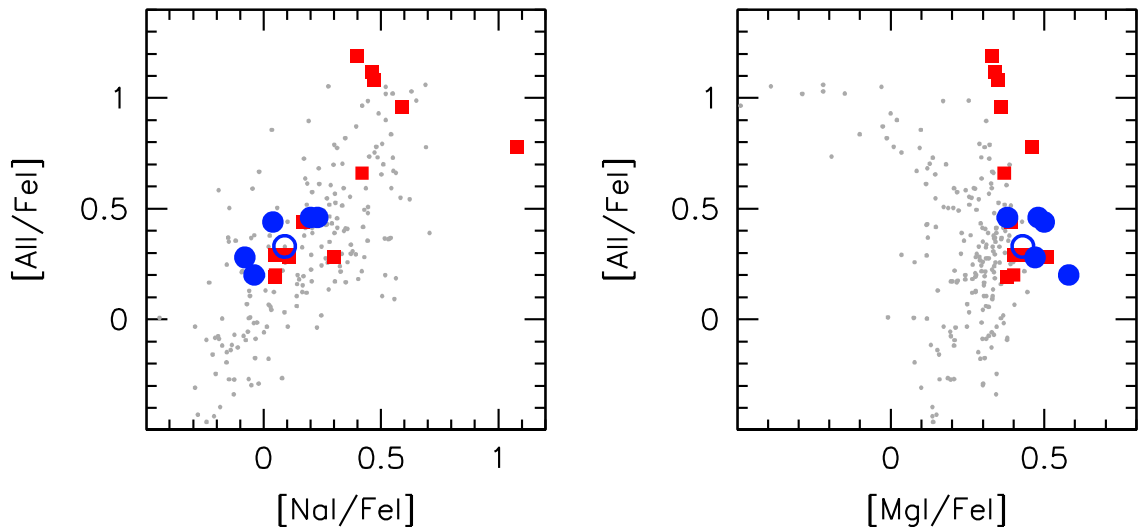}
\caption{Aluminum-sodium correlation (left panel) and
  aluminum-magnesium anti-correlations (right panel) for the observed
  stars (same symbols as in Fig. \ref{light}). Grey dots as in
  Fig. \ref{light}.}
\label{Al}
\end{figure}

For all the studied stars we have also determined the abundances of O,
Na, Al and Mg from the neutral lines available.  As shown in
Figs. \ref{ONa} and \ref{Al}, our sample of RGB stars shows the
typical behaviors observed in all massive GCs, with large and
mutually correlated dispersions of O, Na and Al (and with one
  of the most Na-rich giant ever observed in a GC: RGB star 89, with
  [Na/Fe]=$+1.08$ dex). Instead, the light-element abundances of AGB
stars are essentially constant and clumped at the low-end of the Na
and Al values of the RGB sample.

If the (still unclear) NLTE effects impacting the FeI and TiI
abundances of the AGB targets significantly weaken the
minority species lines (as it seems reasonable to assume), also the
measured abundances of sodium, aluminum and magnesium could be
underestimated for these objects (even when referred to the neutral
iron abundance, FeI). Thus, although the observed
  star-to-star variation of Na in most GCs is often a factor of 5-10
  larger than the suspected NLTE effects on Fe and Ti, we caution that
  it could be risky to derive firm conclusions about a lack of Na-rich
  AGB stars on the basis of the sodium abundance alone until these
  effects are properly understood and quantified (of course, the same
  holds for any other light-element potentially affected by NLTE
  effects, especially if the star-to-star variations of this element
  are intrinsically small). In fact, a lack of Na-rich AGB stars
could be either real, or just due to a bias induced by NLTE effects.
A solid evidence, instead, is obtained if the result is based on
elemental species (like the oxygen abundance derived from the
forbidden line considered here) that are virtually unaffected by NLTE
effects.  Hence, Fig. \ref{ONa}, showing that the oxygen abundances of
all AGB stars are larger than those expected for the SG population and
measured, in fact, for a sub-sample of RGB giants, convincingly
indicates that none of the AGB targets studied in M62 is compatible
with the SG of the cluster.

Does this mean that the SG stars in M62 did not experience the AGB
phase (as it has been suggested for NGC6752 by \citealp{campbell13})?
To answer this question we note that, although variable from cluster
to cluster, the typical percentages of FG and SG stars in Galactic GCs
are 30\% and 70\%, respectively \citep[e.g.][]{carretta13,bastian15b}. On this
basis, we should have observed 4 second generation AGB stars in our
sample. In alternative, from Figs. \ref{ONa} and \ref{Al} we see that
6 out of 13 (46\%) RGB stars likely belong to the SG and, we could
have therefore expected 2-3 AGB stars in the same group,
at odds with what observed.
On the other hand, a deficiency of CN-strong
(second generation) AGB stars in several GCs is known since the
pioneering work of \citet{norris81} and it has been recently found
to be most severe in GCs with the bluest HB morphology
\citep[see, e.g.,][and references therein]{gratton10}.
While M62 has indeed a very extended HB, it shows no deficiency of AGB stars.
In fact, by using ACS and WFC3 HST archive data acquired in the $m_{\rm F390W}$
and $m_{\rm F658N}$ filters, we counted the number of AGB and HB
stars (86 and 640, respectively) in M62, finding that their ratio
(the so-called $R_2$ parameter; \citealp{caputo89}) is $R_2\simeq
0.13$. This value is in very good agreement with the theoretical
predictions based on the ratio between the AGB and the HB
evolutionary timescales \citep[e.g.][]{cassisi01}.
Hence, our observations show that all the sampled AGB stars belong to the FG,
but we cannot exclude that some SG object is present along the AGB of M62.

Clearly, if a complete lack of SG AGB stars is confirmed by future
studies in M62, NGC 6752, M13 \citep[see e.g.][]{sneden00,johnson12} or any other GC,
this will represent a new challenge for the formation and evolution models of these
stellar systems (as already discussed, e.g., by \citealp{charbonnel13}
and \citealp{cassisi14}).

\acknowledgements This research is part of the project Cosmic-Lab (see
http://www.cosmic-lab.eu) funded by the European Research Council
(under contract ERC-2010-AdG-267675).  We warmly thank the anonymous
referee for suggestions that helped improving the paper.


\newpage


\begin{deluxetable}{rccccccrc}
\tablecolumns{9}
\tiny
\tablewidth{0pt}
\tablecaption{Photometric properties and radial velocities of the RGB and AGB sample}
\tablehead{\colhead{ID} & \colhead{R.A.} & \colhead{Decl.} & \colhead{U} & \colhead{V}
& \colhead{U$_{0}$} & \colhead{V$_{0}$} & \colhead{RV} & \colhead{Type} \\
& (J2000) & (J2000) & \colhead{(mag)} & \colhead{(mag)}
& \colhead{(mag)} & \colhead{(mag)} & \colhead{(km s$^{-1}$)} & \colhead{}}
\startdata
  50  &  255.2961736  &  --30.1122536  &  17.458  &  14.061  &  14.558  &  12.184  &  --95.41 $\pm$ 0.06  &  R  \\
  54  &  255.2968276  &  --30.1110148  &  17.465  &  14.149  &  14.462  &  12.206  &  --68.41 $\pm$ 0.04  &  R  \\
  76  &  255.3016326  &  --30.0879873  &  17.578  &  14.416  &  14.734  &  12.576  &  --69.34 $\pm$ 0.06  &  R  \\
  82  &  255.2908040  &  --30.1230200  &  17.378  &  14.478  &  14.649  &  12.712  &  --56.67 $\pm$ 0.04  &  R  \\
  89  &  255.3053120  &  --30.1235390  &  17.694  &  14.584  &  14.759  &  12.685  &  --68.78 $\pm$ 0.04  &  R  \\
  95  &  255.2683150  &  --30.1061800  &  17.689  &  14.624  &  14.821  &  12.768  &  --85.92 $\pm$ 0.06  &  R  \\
  97  &  255.2746210  &  --30.1078150  &  17.551  &  14.632  &  14.703  &  12.789  &  --92.22 $\pm$ 0.06  &  R  \\
 104  &  255.2990264  &  --30.1195799  &  17.502  &  14.680  &  14.861  &  12.971  &  --81.19 $\pm$ 0.04  &  R  \\
 118  &  255.2953240  &  --30.1054710  &  17.584  &  14.771  &  14.883  &  13.023  &  --90.41 $\pm$ 0.06  &  R  \\
 127  &  255.3064600  &  --30.0967810  &  17.775  &  14.895  &  15.020  &  13.112  &  --55.24 $\pm$ 0.06  &  R  \\
 133  &  255.3025803  &  --30.1265560  &  17.819  &  14.939  &  14.903  &  13.052  &  --90.70 $\pm$ 0.05  &  R  \\
 145  &  255.2959190  &  --30.1263240  &  17.831  &  15.041  &  15.031  &  13.229  &  --63.56 $\pm$ 0.05  &  R  \\
 157  &  255.2998135  &  --30.0934941  &  17.720  &  15.174  &  14.988  &  13.406  &  --74.04 $\pm$ 0.05  &  R  \\
\hline
  79  &  255.3060883  &  --30.1031433  &  17.335  &  14.430  &  14.443  &  12.558  & --109.85 $\pm$ 0.06  &  A  \\
  96  &  255.2885360  &  --30.1173880  &  17.345  &  14.629  &  14.558  &  12.826  &  --81.49 $\pm$ 0.07  &  A  \\
 116  &  255.2778880  &  --30.1205350  &  17.130  &  14.764  &  14.348  &  12.963  &  --87.57 $\pm$ 0.09  &  A  \\
 128  &  255.2980470  &  --30.1078870  &  17.248  &  14.895  &  14.501  &  13.117  &  --72.72 $\pm$ 0.08  &  A  \\
 135  &  255.2914560  &  --30.1287900  &  17.416  &  14.952  &  14.613  &  13.138  &  --60.08 $\pm$ 0.07  &  A  \\
 158  &  255.3017290  &  --30.1013070  &  17.361  &  15.180  &  14.647  &  13.424  &  --53.49 $\pm$ 0.07  &  A  \\
\enddata
\tablecomments{Identification number, coordinates, $U$, $V$, $U_{0}$ and $V_{0}$
magnitudes, heliocentric radial velocity, and type of star (R=RGB, A=AGB).}
\label{tab1}
\end{deluxetable}

\begin{deluxetable}{ccrcc}
\tablecolumns{5}
\tablewidth{0pt}
\tablecaption{Wavelength, element, oscillator strength, excitation potential, and reference source of adopted lines.}
\tablehead{\colhead{Wavelength} & \colhead{El.} & \colhead{log $gf$} & \colhead{E.P.} & \colhead{Ref.} \\
\colhead{($\mathring{\rm A}$)} & & & \colhead{(eV)} & }
\startdata
 4962.572  &  FeI  &  --1.182  &  4.178  &  \cite{fuhr06}   \\
 4967.897  &  FeI  &  --0.534  &  4.191  &  K   \\
 4969.917  &  FeI  &  --0.710  &  4.217  &  \cite{fuhr88}   \\
 4982.499  &  FeI  &    0.164  &  4.103  &  K   \\
 4983.250  &  FeI  &  --0.111  &  4.154  &  K   \\
 4985.547  &  FeI  &  --1.331  &  2.865  &  \cite{fuhr06}   \\
 4950.106  &  FeI  &  --1.670  &  3.417  &  \cite{fuhr88}   \\
 4962.572  &  FeI  &  --1.182  &  4.178  &  \cite{fuhr06}   \\
 4967.897  &  FeI  &  --0.534  &  4.191  &  K   \\
 4969.917  &  FeI  &  --0.710  &  4.217  &  \cite{fuhr88}   \\
\enddata
\tablecomments{
K = Oscillator strengths (OS) from the R.L.Kurucz on-line database of observed and 
predicted atomic transitions (see http://kurucz.harvard.edu),
NIST = OS from NIST database (see http://www.nist.gov/pml/data/asd.cfm)
S = OS from solar analysis by F. Castelli (see http://wwwuser.oats.inaf.it/castelli/linelists.html).
For AlI lines we derived astrophysical oscillator strengths (labeled as S*) by using the solar flux spectra of
\cite{neckel84} and the model atmosphere for the Sun computed by
F. Castelli\footnote{http://wwwuser.oats.inaf.it/castelli/sun/ap00t5777g44377k1odfnew.dat}
adopting the solar abundances of \cite{grevesse98}.
The entire Table is available in the on-line version, a portion
is shown here for guidance about its form and content.
}
\label{tab2}
\end{deluxetable}

\newpage

\begin{landscape}
\begin{deluxetable}{rcccrcrcrcrc}
\tablecolumns{12}
\scriptsize
\tablewidth{0pt}
\tablecaption{Atmospheric parameters, iron and titanium abundances of the measured RGB and AGB stars.}
\tablehead{\colhead{ID} & \colhead{$T_{\rm eff}$} & \colhead{log g}
&  \colhead{$v_{\rm turb}$}
& \colhead{[FeI/H]} & \colhead{n$_{\rm FeI}$} & \colhead{[FeII/H]} & \colhead{n$_{\rm FeII}$} 
& \colhead{[TiI/H]} & \colhead{n$_{\rm TiI}$} & \colhead{[TiII/H]} & \colhead{n$_{\rm TiII}$} \\
& \colhead{(K)} & \colhead{(dex)} &  \colhead{(km s$^{-1}$)} & \colhead{(dex)} & & \colhead{(dex)} & 
& \colhead{(dex)} & & \colhead{(dex)} & }
\startdata
  50  &  4225  &  0.85  &  1.30  &  --1.13 $\pm$ 0.01  &  128  &  --1.13 $\pm$ 0.02  &  12  &  --0.91 $\pm$ 0.01  &  58  & --0.95 $\pm$ 0.05  &  12  \\
  54  &  4215  &  0.85  &  1.40  &  --1.17 $\pm$ 0.01  &  130  &  --1.14 $\pm$ 0.01  &   7  &  --0.99 $\pm$ 0.01  &  63  & --1.06 $\pm$ 0.03  &  14  \\
  76  &  4375  &  1.15  &  1.35  &  --1.05 $\pm$ 0.01  &  106  &  --1.00 $\pm$ 0.03  &   7  &  --0.87 $\pm$ 0.02  &  37  & --0.88 $\pm$ 0.05  &   6  \\
  82  &  4295  &  1.15  &  1.30  &  --1.06 $\pm$ 0.01  &  104  &  --1.02 $\pm$ 0.03  &  11  &  --0.89 $\pm$ 0.01  &  45  & --0.92 $\pm$ 0.06  &   7  \\
  89  &  4355  &  1.15  &  1.50  &  --1.07 $\pm$ 0.01  &  127  &  --1.08 $\pm$ 0.03  &  10  &  --0.78 $\pm$ 0.01  &  62  & --0.93 $\pm$ 0.03  &  14  \\
  95  &  4365  &  1.20  &  1.45  &  --1.07 $\pm$ 0.01  &  134  &  --1.10 $\pm$ 0.02  &  11  &  --0.89 $\pm$ 0.01  &  58  & --0.93 $\pm$ 0.04  &  15  \\
  97  &  4425  &  1.25  &  1.40  &  --1.01 $\pm$ 0.01  &  142  &  --1.02 $\pm$ 0.02  &  12  &  --0.82 $\pm$ 0.01  &  50  & --0.96 $\pm$ 0.05  &  14  \\
 104  &  4325  &  1.20  &  1.30  &  --1.11 $\pm$ 0.01  &  108  &  --1.00 $\pm$ 0.03  &   7  &  --0.94 $\pm$ 0.01  &  40  & --0.90 $\pm$ 0.05  &   7  \\
 118  &  4450  &  1.35  &  1.40  &  --1.05 $\pm$ 0.01  &  140  &  --1.03 $\pm$ 0.01  &   8  &  --0.81 $\pm$ 0.01  &  56  & --0.88 $\pm$ 0.03  &  13  \\
 127  &  4425  &  1.35  &  1.35  &  --1.06 $\pm$ 0.01  &  102  &  --0.95 $\pm$ 0.02  &  10  &  --0.90 $\pm$ 0.02  &  57  & --0.90 $\pm$ 0.05  &  15  \\
 133  &  4450  &  1.35  &  1.40  &  --1.10 $\pm$ 0.01  &  142  &  --1.05 $\pm$ 0.01  &   9  &  --0.89 $\pm$ 0.01  &  57  & --0.91 $\pm$ 0.04  &  15  \\
 145  &  4475  &  1.45  &  1.30  &  --1.06 $\pm$ 0.01  &  146  &  --0.98 $\pm$ 0.02  &  10  &  --0.88 $\pm$ 0.01  &  47  & --0.92 $\pm$ 0.04  &  13  \\
 157  &  4545  &  1.55  &  1.45  &  --1.04 $\pm$ 0.01  &  136  &  --1.01 $\pm$ 0.02  &  10  &  --0.83 $\pm$ 0.01  &  52  & --0.84 $\pm$ 0.05  &  13  \\
\hline
  79  &  4415  &  1.00  &  1.55  &  --1.19 $\pm$ 0.01  &  131  &  --1.08 $\pm$ 0.01  &   8  &  --1.08 $\pm$ 0.01  &  48  & --1.03 $\pm$ 0.04  &  15  \\
  96  &  4450  &  1.15  &  1.50  &  --1.10 $\pm$ 0.01  &  130  &  --1.13 $\pm$ 0.03  &  11  &  --0.71 $\pm$ 0.03  &  33  & --0.88 $\pm$ 0.07  &   9  \\
 116  &  4760  &  1.35  &  1.80  &  --1.24 $\pm$ 0.01  &  134  &  --1.13 $\pm$ 0.03  &   9  &  --1.12 $\pm$ 0.02  &  27  & --0.99 $\pm$ 0.04  &  13  \\
 128  &  4760  &  1.45  &  1.60  &  --1.21 $\pm$ 0.01  &  138  &  --1.10 $\pm$ 0.04  &  12  &  --1.11 $\pm$ 0.02  &  33  & --0.95 $\pm$ 0.04  &  14  \\
 135  &  4635  &  1.40  &  1.55  &  --1.14 $\pm$ 0.01  &  128  &  --0.98 $\pm$ 0.03  &  10  &  --1.08 $\pm$ 0.02  &  33  & --0.88 $\pm$ 0.04  &  13  \\
 158  &  4840  &  1.60  &  1.65  &  --1.19 $\pm$ 0.01  &  142  &  --1.01 $\pm$ 0.02  &  11  &  --1.10 $\pm$ 0.02  &  27  & --0.92 $\pm$ 0.05  &  14  \\
\enddata
\tablecomments{Identification number, spectroscopic temperature and
  photometric gravities, microturbulent velocities, iron and titanium
  abundances with internal uncertainty and number of used lines, as
  measured from neutral and single-ionized lines. For all the stars a
  global metallicity of [M/H]$ = -1.0$ dex has been assumed for the
  model atmosphere. The adopted solar values are from \cite{grevesse98}.}
\label{tab3}
\end{deluxetable}
\end{landscape}

\begin{landscape}
\begin{deluxetable}{rcrrccccccccc}
\tablecolumns{13}
\scriptsize
\tablewidth{0pt}
\tablecaption{OI, NaI, MgI, AlI, TiI and TiII abundances of the RGB and AGB sample}
\tablehead{\colhead{ID} & \colhead{[OI/FeII]} & \colhead{[NaI/FeI]$_{\rm LTE}$} & \colhead{[NaI/FeI]$_{\rm NLTE}$}
& \colhead{n$_{\rm Na}$} & \colhead{[MgI/FeI]} & \colhead{n$_{\rm Mg}$} & \colhead{[AlI/FeI]} & \colhead{n$_{\rm Al}$}
& \colhead{[TiI/FeI]} & \colhead{n$_{\rm TiI}$} & \colhead{[TiII/FeII]} & \colhead{n$_{\rm TiII}$}\\
& \colhead{(dex)} & \colhead{(dex)} & \colhead{(dex)} & & \colhead{(dex)} & & \colhead{(dex)} & &
\colhead{(dex)} & & \colhead{(dex)} & }
\startdata
  50  &  0.39 $\pm$ 0.05  & --0.01 $\pm$ 0.12  &   0.11 $\pm$ 0.09  &  4  &  0.47 $\pm$ 0.03  &  3  &  0.28 $\pm$ 0.01  &  2  &  0.22 $\pm$ 0.02  &  58  &  0.18 $\pm$ 0.05  &  12  \\
  54  &  0.35 $\pm$ 0.04  &   0.17 $\pm$ 0.10  &   0.30 $\pm$ 0.06  &  4  &  0.51 $\pm$ 0.02  &  3  &  0.28 $\pm$ 0.00  &  2  &  0.18 $\pm$ 0.01  &  63  &  0.08 $\pm$ 0.03  &  14  \\
  76  &       $<$ --0.16  &   0.34 $\pm$ 0.11  &   0.42 $\pm$ 0.07  &  4  &  0.37 $\pm$ 0.02  &  3  &  0.66 $\pm$ 0.02  &  2  &  0.18 $\pm$ 0.02  &  37  &  0.12 $\pm$ 0.05  &   6  \\
  82  &  0.31 $\pm$ 0.07  & --0.05 $\pm$ 0.08  &   0.05 $\pm$ 0.04  &  4  &  0.40 $\pm$ 0.03  &  3  &  0.20 $\pm$ 0.02  &  2  &  0.17 $\pm$ 0.02  &  45  &  0.09 $\pm$ 0.07  &   7  \\
  89  &       $<$ --0.28  &   1.04 $\pm$ 0.06  &   1.08 $\pm$ 0.02  &  4  &  0.46 $\pm$ 0.02  &  3  &  0.78 $\pm$ 0.04  &  2  &  0.29 $\pm$ 0.02  &  62  &  0.15 $\pm$ 0.04  &  14  \\
  95  &       $<$ --0.36  &   0.31 $\pm$ 0.17  &   0.40 $\pm$ 0.16  &  4  &  0.33 $\pm$ 0.05  &  3  &  1.19 $\pm$ 0.06  &  2  &  0.18 $\pm$ 0.02  &  58  &  0.17 $\pm$ 0.05  &  15  \\
  97  &       $<$ --0.34  &   0.41 $\pm$ 0.14  &   0.47 $\pm$ 0.09  &  4  &  0.35 $\pm$ 0.06  &  2  &  1.08 $\pm$ 0.04  &  2  &  0.19 $\pm$ 0.02  &  50  &  0.06 $\pm$ 0.05  &  14  \\
 104  &  0.25 $\pm$ 0.08  &   0.01 $\pm$ 0.11  &   0.11 $\pm$ 0.07  &  4  &  0.43 $\pm$ 0.06  &  3  &  0.29 $\pm$ 0.05  &  1  &  0.17 $\pm$ 0.02  &  40  &  0.10 $\pm$ 0.06  &   7  \\
 118  &  0.39 $\pm$ 0.05  & --0.03 $\pm$ 0.10  &   0.05 $\pm$ 0.07  &  4  &  0.40 $\pm$ 0.06  &  2  &  0.29 $\pm$ 0.03  &  2  &  0.24 $\pm$ 0.02  &  56  &  0.15 $\pm$ 0.03  &  13  \\
 127  &  0.05 $\pm$ 0.09  &   0.08 $\pm$ 0.11  &   0.17 $\pm$ 0.07  &  4  &  0.39 $\pm$ 0.02  &  3  &  0.44 $\pm$ 0.04  &  1  &  0.16 $\pm$ 0.02  &  57  &  0.05 $\pm$ 0.05  &  15  \\
 133  &       $<$ --0.17  &   0.40 $\pm$ 0.10  &   0.46 $\pm$ 0.06  &  4  &  0.34 $\pm$ 0.05  &  2  &  1.12 $\pm$ 0.06  &  2  &  0.21 $\pm$ 0.01  &  57  &  0.14 $\pm$ 0.04  &  15  \\
 145  &  0.13 $\pm$ 0.07  & --0.03 $\pm$ 0.08  &   0.05 $\pm$ 0.06  &  4  &  0.38 $\pm$ 0.01  &  3  &  0.19 $\pm$ 0.04  &  2  &  0.18 $\pm$ 0.01  &  47  &  0.07 $\pm$ 0.04  &  13  \\
 157  &       $<$ --0.40  &   0.55 $\pm$ 0.10  &   0.59 $\pm$ 0.06  &  4  &  0.36 $\pm$ 0.02  &  3  &  0.96 $\pm$ 0.03  &  2  &  0.21 $\pm$ 0.01  &  52  &  0.17 $\pm$ 0.05  &  13  \\
\hline
  79  &  0.37 $\pm$ 0.05  & --0.20 $\pm$ 0.06  & --0.04 $\pm$ 0.04  &  4  &  0.58 $\pm$ 0.08  &  2  &  0.20 $\pm$ 0.07  &  1  &  0.11 $\pm$ 0.01  &  48  &  0.05 $\pm$ 0.04  &  15  \\
  96  &  0.35 $\pm$ 0.06  & --0.03 $\pm$ 0.08  &   0.09 $\pm$ 0.04  &  4  &  0.43 $\pm$ 0.05  &  3  &  0.33 $\pm$ 0.06  &  1  &  0.39 $\pm$ 0.04  &  33  &  0.25 $\pm$ 0.08  &   9  \\
 116  &  0.17 $\pm$ 0.07  &   0.13 $\pm$ 0.06  &   0.23 $\pm$ 0.03  &  4  &  0.38 $\pm$ 0.04  &  2  &  0.46 $\pm$ 0.05  &  1  &  0.12 $\pm$ 0.02  &  27  &  0.14 $\pm$ 0.05  &  13  \\
 128  &  0.19 $\pm$ 0.06  & --0.06 $\pm$ 0.10  &   0.04 $\pm$ 0.09  &  4  &  0.50 $\pm$ 0.09  &  3  &  0.44 $\pm$ 0.07  &  1  &  0.10 $\pm$ 0.03  &  33  &  0.15 $\pm$ 0.06  &  14  \\
 135  &  0.30 $\pm$ 0.07  & --0.20 $\pm$ 0.04  & --0.08 $\pm$ 0.03  &  4  &  0.47 $\pm$ 0.03  &  3  &  0.28 $\pm$ 0.07  &  1  &  0.06 $\pm$ 0.02  &  33  &  0.11 $\pm$ 0.05  &  13  \\
 158  &  0.19 $\pm$ 0.05  &   0.13 $\pm$ 0.06  &   0.20 $\pm$ 0.03  &  3  &  0.48 $\pm$ 0.08  &  3  &  0.46 $\pm$ 0.05  &  1  &  0.09 $\pm$ 0.02  &  27  &  0.09 $\pm$ 0.05  &  14  \\
\hline
\enddata
\tablecomments{The oxygen abundance has been derived from the 6300.3$\rm\mathring{A}$ [OI] line,
the abundances of sodium have been reported without and with NLTE corrections computed following
\cite{gratton99}.
The reference solar values are taken from \cite{caffau11} for the oxygen,
from \cite{grevesse98} for the other species.}
\label{tab4}
\end{deluxetable}
\end{landscape}

\begin{deluxetable}{ccccc}
\tablecolumns{5}
\tiny
\tablewidth{0pt}
\tablecaption{Abundance uncertainties due to the atmospheric parameters for the stars 157 and 158.}
\tablehead{\colhead{Species} & \colhead{Global} & \colhead{$\delta T_{\rm eff}$}
& \colhead{$\delta \log g$} & \colhead{$\delta v_{\rm turb}$} \\
\colhead{} & \colhead{Uncertainty} & \colhead{$\pm 50 K$}
& \colhead{$\pm 0.1$} & \colhead{$\pm 0.1 km s^{-1}$} \\
& \colhead{(dex)} & \colhead{(dex)} & \colhead{(dex)} & \colhead{(dex)}}
\startdata
& & 157 (RGB) & & \\
\hline
 FeI  &  $\pm$0.07  &  $\pm$0.04  &  $\pm$0.00  &  $\mp$0.06  \\
 FeII &  $\pm$0.08  &  $\mp$0.05  &  $\pm$0.05  &  $\mp$0.04  \\
 OI   &  $\pm$0.04  &  $\pm$0.01  &  $\pm$0.03  &  $\mp$0.02  \\
 NaI  &  $\pm$0.05  &  $\pm$0.04  &  $\mp$0.01  &  $\mp$0.02  \\
 MgI  &  $\pm$0.04  &  $\pm$0.03  &  $\pm$0.00  &  $\mp$0.03  \\
 AlI  &  $\pm$0.04  &  $\pm$0.04  &  $\pm$0.00  &  $\mp$0.02  \\
 TiI  &  $\pm$0.09  &  $\pm$0.08  &  $\pm$0.00  &  $\mp$0.03  \\
 TiII &  $\pm$0.05  &  $\mp$0.02  &  $\pm$0.04  &  $\mp$0.03  \\
\hline
& & 158 (AGB) & & \\
\hline
 FeI  &  $\pm$0.07  &  $\pm$0.06  &  $\pm$0.00  &  $\mp$0.04  \\
 FeII &  $\pm$0.07  &  $\mp$0.03  &  $\pm$0.05  &  $\mp$0.03  \\
 OI   &  $\pm$0.05  &  $\pm$0.02  &  $\pm$0.04  &  $\mp$0.02  \\
 NaI  &  $\pm$0.04  &  $\pm$0.04  &  $\pm$0.00  &  $\mp$0.01  \\
 MgI  &  $\pm$0.03  &  $\pm$0.03  &  $\pm$0.00  &  $\mp$0.01  \\
 AlI  &  $\pm$0.03  &  $\pm$0.03  &  $\pm$0.00  &  $\mp$0.00  \\
 TiI  &  $\pm$0.08  &  $\pm$0.08  &  $\pm$0.00  &  $\mp$0.01  \\
 TiII &  $\pm$0.06  &  $\mp$0.01  &  $\pm$0.05  &  $\mp$0.03  \\
\hline
\enddata
\tablecomments{The second column shows the global uncertainty
  calculated by adding in quadrature the single uncertainties.  The
  other columns list the uncertainties obtained by varying only one
  parameter at a time, while keeping the others fixed.}
\label{tab5}
\end{deluxetable}


\end{document}